# Towards wafer scale inductive determination of magnetostatic and dynamic parameters of magnetic thin films and multilayers

S. Sievers[1], N. Liebing[1], P. Nass[1], S. Serrano-Guisan[1], M. Pasquale[2], H.W. Schumacher[1]

[1]Physikalisch-Technische Bundesanstalt, Bundesalle 100, D-38116 Braunschweig, Germany
[2]Instituto Nazionale di Ricerca Metrologica, Strada della Cacce 91, 10135 Torino, Italy

We investigate an inductive probe head suitable for non-invasive characterization of the magnetostatic and dynamic parameters of magnetic thin films and multilayers on the wafer scale. The probe is based on a planar waveguide with rearward high frequency connectors that can be brought in close contact to the wafer surface. Inductive characterization of the magnetic material is carried out by vector network analyzer ferromagnetic resonance. Analysis of the field dispersion of the resonance allows the determination of key material parameters such as the saturation magnetization $M_S$ or the effective damping parameter $\alpha_{eff}$. Three waveguide designs are tested. The broadband frequency response is characterized and the suitability for inductive determination of $M_S$ and $\alpha_{eff}$ is compared. Integration of such probes in a wafer prober could in the future allow wafer scale in-line testing of magnetostatic and dynamic key material parameters of magnetic thin films and multilayers.

I. INTRODUCTION

The development of fast, reliable and non-invasive characterization tools is the key issue for the development and production of thin film based devices. For magnetic thin films and multilayers magneto static key material parameters are e.g. the saturation magnetization $M_S$, the interlayer couplings J and the different anisotropy contributions $H_{Ki}$. For high frequency applications also the effective magnetic damping $\alpha$ can be of great importance as it determines the decay of magnetization precession. Recent developments in the field of spin torque magnetic random access memories (ST-MRAM) have further increased the interest in characterizing dissipative processes in magnetic systems. Indeed, in the Slonzcewski model the critical current density $j_c$ for ST magnetization reversal [7] is directly proportional to $\alpha$. According to this model also other material parameters such as the saturation magnetization $M_S$ enter in the relation for $j_c$. While $j_C$ presently can only be derived by a combination of highly elaborated nanofabrication and consecutive electrical measurements of individual sub-micron magnetic tunnel junction (MTJ) pillars, $\alpha$ and $M_S$ of thin films and multilayers can be measured non invasively e.g. by inductive techniques based on ferromagnetic resonance [1-5]. This might pave the way to a future non invasive characterization of $j_C$ of spin torque material stacks based on the inductively determined values of $\alpha$ and $M_S$.

For industrial applications, a measurement of $\alpha$ on integral wafers, preferentially in-line, is requested. However, the conventional way to perform inductive measurements on magnetic films is either by cavity based ferromagnetic resonance (FMR) [8] of millimeter sized samples or by placing a millimetre-sized sample on the center area of a coplanar wave guide line for pulsed inductive microwave magentometry (PIMM) [5] or for vector network analyzer ferromagnetic resonance (VNA)-FMR. For the latter the wave guide is connected to a VNA either by impedance matched microwave probes or by suitable end launch connectors. However both of these wave guide and connector geometries limit the sample size to dimensions smaller than the size of the wave guide thereby inhibiting wafer scale inductive testing.

In this manuscript we investigate an inductive probe head which overcomes the above limitations and might in the future allow wafer scale inductive testing of magnetic thin films and multilayers. The inductive probe head is based on a planar transmission line with rearward high frequency connectors. The probe head shows good high frequency properties allowing VNA-FMR measurements of magnetic thin films. Three different transmission line geometries are compared and tested by VNA-FMR measurements of a soft magnetic thin film. The influence of the sample size on the derived magnetostatic and dynamic parameters results is discussed. The derived results are comparable to data obtained by a conventional coplanar FMR probe on the same films.

## II. EXPERIMENTAL SETUP

A probe head for the inductive characterization of magnetic thin films on the wafer scale has to fulfill the following requirements:

I) The geometrical requirements for a wafer prober require rear side high frequency contacts as discussed above.

II) The transmission line should be well matched to an impedance of 50 Ω. It should show a frequency independent transmission in the relevant FMR frequency range or should at least allow a well defined normalization of the transmission in that frequency range. Furthermore the transmission should not significantly depend on the applied magnetic field.

III) The transmission line design must allow effective excitation and detection of uniform magnetization precession in the magnetic thin film.

Photographs of the three different transmission line geometries are found in Fig. 1 (a)-(c). Two probe head types (a), (b) use coplanar waveguides with a back ground plane while probe head (c) uses a microstrip design again with a back ground plane. Types (a) and (b) differ in that type (b) has four additional through connector holes between the top and back ground metallization to suppress internal resonances [6]. The waveguides were fabricated on 1.5mm thick FR04 printed circuit board (PCB). The wave guide layout is numerically optimized to allow 50 Ω impedance matching. The gold metallization for the coplanar line and the back metallization was applied without Ni adhesive coating to avoid spurious magnetic signals. The transmission lines are contacted from the back through metalized mounting holes via soldered SMA connectors with 18 GHz bandwidth as shown in Fig. 1(d).

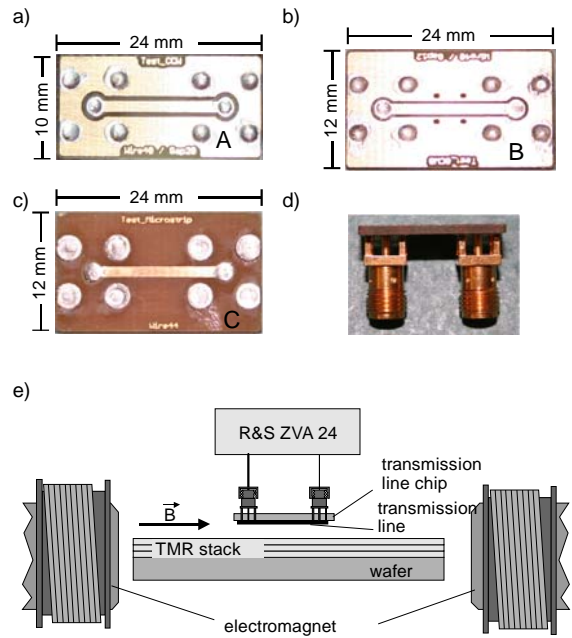

Fig. 1. (a-c) Photos of the different transmission lines layouts. (d) Side view of probe head (a) with mounted rearward SMA connectors. (e) Schematic drawing of the experimental setup for wafer probing. The transmission line chip with the transmission line can be positioned at any point of the sample surface. The sample and the probe head are placed between the coils of an electromagnet generating the external field. FMR measurements are performed using a commercial VNA.

The inductive measurement setup used in our experiment is sketched in Fig. 1(e). The planar probe head is placed on top of the wafer in direct contact to the magnetic thin film under investigation. The probe head is mounted between the pole shoes of an electromagnet allowing the application of homogeneous static fields up to B = 1 T. In the present setup the maximum wafer size is limited by the pole shoe separation to 7 cm. The probe head can be positioned at different positions on the wafer to investigate the magnetic properties at different positions of the sample enabling future wafer scale scanning. 18 GHz bandwidth cables connect the SMA connectors of the probe head to a commercial vector network analyzer (Rhode & Schartz ZVA 24).

## III. RESULTS

Initially the three different probe heads were characterized by VNA stray parameter measurements without magnetic sample. Fig. 2 shows typical measurements of the three different probe heads (a)-(c). Here the insertion loss (IL) is plotted as function of frequency $f$. Before the measurement a transmission normalization through the complete assembly building the wafer prober consisting of the semi rigid connection cables the connectors and the transmission line was used to calibrate the system for the determination of the S21 stray parameter. S21 is the ratio of the incoming to the outgoing signals. The IL is determined from S21 as IL = -20·log$_{10}$|S21|.

The three bottom plots show the normalized IL for heads (a)-(c) at a low field of 10 mT. The curves are offset by 0.5 dB for clarity. All three heads show a flat response of the normalized IL. Up to 17 GHz IL varies below 0±0.1dB and below 0±0.3dB up to 18 GHz. These fluctuations can be attributed to internal resonances in the transmission line and connectors. Also the influence of the applied field B has been tested. The upper curve exemplarily shows a measurement of the transmission head (a) at B = 300 mT. Though slight changes of the transmission properties upon field application are present these are again limited to below 0±0.1dB. The other transmission lines (b),(c) show a similar behavior in field (not shown) in the given frequency range. These heads thus fulfill the criterion (II) and should hence be suitable for inductive FMR measurements on the wafer scale.

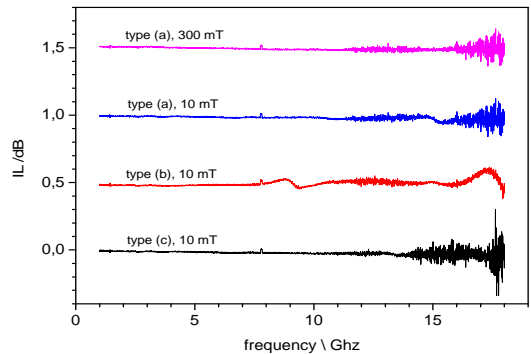

Fig. 2. Insertion loss IL of the three probe heads (a)-(c) at a residual field of 10 mT. The curves are offset by 0.5 dB for clarity. In the frequency range from 1 to 18 GHz all probe heads weak deviations (±0.1 dB) from constant normalized transmission. The topmost curve shows the IL of proba (a) after increasing the applied field to 300 mT. No significant magnetic field dependence of the transmission is found.

VNA-FMR measurements using the three heads were carried out on a 10 nm thick sputter deposited $Co_{66}Fe_{22}B_{12}$ thin film which has been previously characterized by inductive measurements using micropatterned coplanar waveguides [9]. The thin film was sputter deposited in a Singulus Cluster Tool on a not intentionally doped Si wafer.

Standard VNA-FMR measurements are usually carried out on mm sized samples and not on wafers with larger lateral dimensions than the wave guide [1-5]. To test the influence of the sample size on the measurements and the derived parameters the wafer was cut into samples of different sizes from 2 mm x 2 mm to 24 mm x 12 mm. For VNA-FMR the samples with different edge lengths X,Y were centered under the wave guide. The samples were further placed in two orientations with the long dimension either parallel or perpendicular to the applied field B.

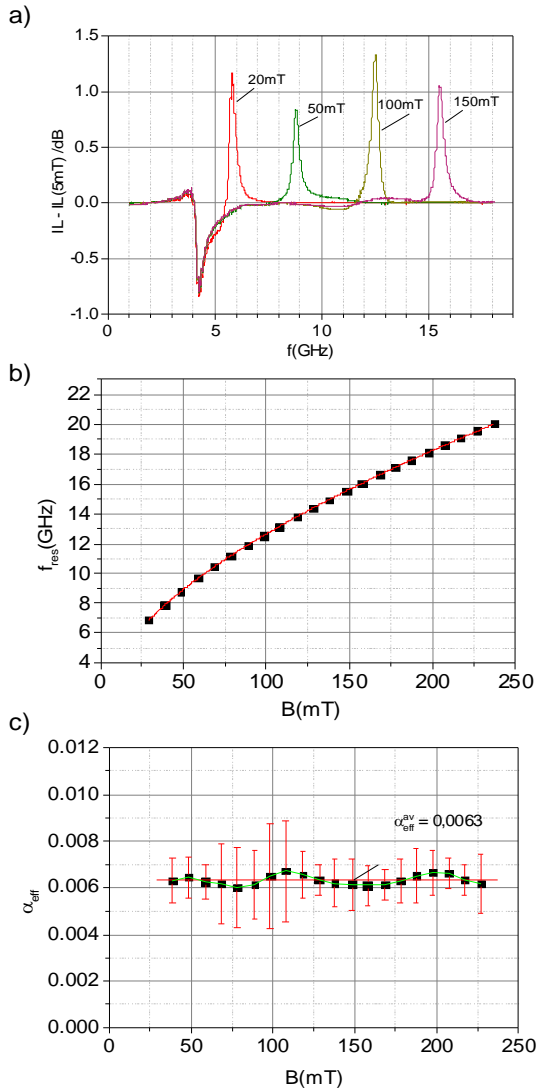

Fig. 3. Results of the VNA-FMR measurements on a CoFeB sample with a size of 14×4 mm² using probe head (a). a) Resonance curves for different fields B after reference subtraction. b) Resonance frequency as a function of the applied field, the red line shows the Kittel fit. c) Resulting values of $\alpha_{eff}$ as a function of the applied field. The $\alpha_{eff}$ data can be well described by a constant average value $\alpha_{eff}^{av}$

For all samples VNA-FMR measurements were performed in the field range from 10 mT to 300 mT. The transmission stray parameters were measured in frequency domain over a range from 1 to 18 GHz at 0 dBm power (1 mW) and with 10 kHz bandwidth. To separate the FMR contributions to the IL from the contributions of e.g. the dielectric material of the sample a reference measurement of S21 and $IL_{ref}$ was carried out at a low field of $B_{ref}$ = 5 mT. Then IL at a given measurement field B was determined and $IL_{ref}$ was subtracted.

Fig. 3(a) shows a typical measurement of the such derived VNA-FMR signal for four different applied field values B = 20, 50, 100, 150 mT. The data is taken using head (a) on a 14×4 mm² sample with the long axis oriented parallel to B. At about 4.2 GHz all measurements show a dip. This dip can be assigned to FMR at the reference field $B_{ref}$ and results from reference subtraction. Furthermore for each field value a well defined positive resonance peak is visible. These peaks have amplitudes of about 1 dB and are hence significantly larger than the IL fluctuations of the probe heads. As expected for FMR the peaks shift to higher frequencies with increasing applied field.

For all field values the FMR peak in the IL data was fitted by a Lorentzian function with amplitude A and the offset $IL_0$

$$IL = IL_0 + \frac{A}{2\pi} \frac{\Delta f / 2}{(f - f_{FMR})^2 + (\Delta f / 2)^2}$$

to determine the center frequency $f_{FMR}$ and the line width $\Delta f$ of the FMR.

Fig. 3b shows the field dependence of $f_{FMR}$ for all measured field values. The measured data (black squares) shows the typical FMR behavior that can be well described by a simple Kittel model with the saturation magnetization $M_S$ as the fit parameter (red line). For the given data the fit yields a value of $\mu_0 M_S = 1.86$ T.

In addition to $M_S$ also the damping can be derived from the inductive data. $\alpha_{eff}$ can be calculated from $\Delta f$ via [3]

$$\alpha_{eff} = \frac{2\pi \cdot \Delta f}{\gamma \mu_0 (2H_{eff} + M_S)}.$$

Here, $\gamma$ is the free electron gyromagnetic ratio, $\mu_0$ is the vacuum permeability, and $\mu_0 H_{eff} = B$ is the applied field when neglecting anisotropy contributions. Fig. 3(c) shows $\alpha_{eff}$ as function of B. The error bars are derived from the uncertainty of the Lorentzian fit. $\alpha_{eff}$ shows no significant field dependence and the value strays around a constant average value $\alpha_{eff}^{av} = 0.0063 \pm 0.0005$.

Similar measurements and analysis has been carried out for the other sample sizes and on the other two probe heads (b) and (c). For probe head (b) similar results as in Fig. 3 have been obtained. In contrast the FMR data of head (c) shows only very weakly developed FMR peaks (both not shown). Here a meaningful FMR analysis could only carried out for the largest samples. For this strip line design the magnetic field generated by the central conductor is concentrated in the PCB substrate. Therefore FMR excitation and detection is less efficient and criterion III) is not well fulfilled. In contrast the two heads with coplanar design (a) and (b) fulfill all three criteria for FMR wafer probing.

The derived values of $M_S$ and $\alpha_{eff}$ of all samples and probe heads are compiled in Fig. 4. Fig. 4(a) shows $M_S$ derived for all probe heads and samples. $M_S$ strays around an average value of $\mu_0 M_S = 1.86 \pm 0.06$ T. The stray might be related to additional anisotropy contributions such as the varying shape anisotropy of the differently sized samples which have not been taken into account into the Kittel evaluation of the FMR data.

In previous work [9] a 5×5 mm² sample of the same wafer had been inductively characterized by PIMM measurements on micropatterned coplanar waveguides [10].

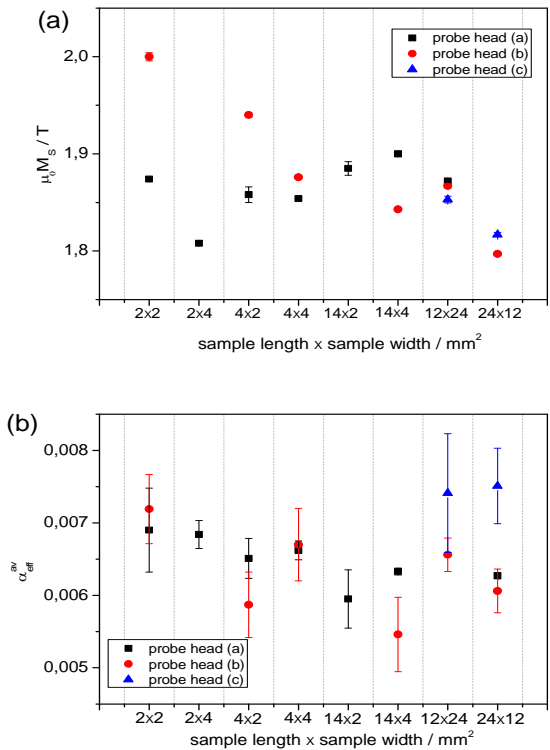

Fig. 4. Effective saturation magnetization $\mu_0 M_S$ (a) and average effective damping parameter $\alpha_{eff}^{av}$ (b) as a function of the sample size and orientation for the three different probe head designs. For probe head (c) only the data for the largest sample could be evaluated.

Within the given uncertainties the previously derived saturation magnetization of $M_S$ = 1.79±0.03 T agrees with the average value of $M_S$ derived by the new inductive probe heads.

The derived values of the effective damping $\alpha_{eff}^{av}$ are plotted in Fig. 4(b). The values vary between 0.0055 and 0.0072 with a mean value of $\alpha_{eff}^{av}$ = 0.0066±0.0006. Except for the values derived by probe head (c) (blue triangles) the derived damping values shows a slight tendency to decrease with increasing sample size. Since all samples are of millimetre size, one would not expect a direct influence of the sample size on the physical dissipation of the thin film and thus on $\alpha_{eff}$. However, note that for larger samples the probed sample volume is significantly larger resulting in a better signal to noise ratio of the measurements. Consequently also a more reliable Lorentzian fit can be expected for these samples. Also the mean value of $\alpha_{eff}^{av}$ agrees within the measurement uncertainty with the value of α = 0.008±0.001 previously determined by PIMM measurements. Note that the previous PIMM experiments have been carried out in a different setup under applied static fields in the range of 1 to 18 mT. A similar increase of the effective damping for lower fields was also found by other groups in previous FMR experiments on Permalloy [3].

## IV. CONCLUSIONS

We have designed and characterized three probe head prototypes for VNA-FMR measurements of magnetic thin films and multilayers on the wafer scale. The two coplanar waveguide based probe heads were found suitable for FMR characterization of magnetic thin films in the frequency range of 1-18 GHz. From the VNA-FMR data the saturation magnetization $M_S$ and the effective Gilbert damping $\alpha_{eff}$ have been derived. The such derived values are in agreement with previously derived data obtained by standard PIMM measurements. The presented coplanar probe heads are thus capable of non-invasive probing of key magneto static and dynamic material parameters of magnetic thin films and multilayers on the wafer scale. In particular the determination of α could in the future enable wafer scale non-invasive determination of the spin transfer torque critical current density $j_C$ of spin torque materials such as magnetic tunnel junction stacks.

ACKNOWLEDGMENT

This work has been developed under the EMRP JRP IND 08 MetMags, jointly funded by EU and EMRP participating countries within EURAMET. We thank J. Langer and B. Ocker from Singulus for providing the CoFeB samples.